\documentclass[twocolumn,showpacs,superscriptaddress,aps,prb]{revtex4}

\usepackage{graphicx}

\begin{document}

\title{Enhancement of long-range magnetic order by magnetic field in
      superconducting La$_2$CuO$_{4+y}$}
\author{B. Khaykovich}
\affiliation{Department of Physics and Center for Material
     Science and Engineering, Massachusetts Institute of Technology,
     Cambridge, MA 02139}
\author{Y. S. Lee}
\altaffiliation[Presently at ]{Department of Physics and Center
     for Material Science and Engineering, Massachusetts Institute
     of Technology, Cambridge, MA 02139.}
\author{R. W. Erwin}
\author{S.-H. Lee}
\affiliation{Center for Neutron Research, NIST, Gaithersburg,
     MD 20899-8562}
\author{S. Wakimoto}
\affiliation{Department of Physics, University of Toronto,
     Toronto, Ontario M5S 1A7, Canada}
\author{K. J. Thomas}
\author{M. A. Kastner}
\affiliation{Department of Physics and Center for Material
     Science and Engineering, Massachusetts Institute of Technology,
     Cambridge, MA 02139}
\author{R. J. Birgeneau}
\affiliation{Department of Physics, University of Toronto,
     Toronto, Ontario M5S 1A7, Canada}
\affiliation{Department of Physics and Center for Material
     Science and Engineering, Massachusetts Institute of Technology,
     Cambridge, MA 02139}

\date{\today}

\begin{abstract}
   We report a detailed study, using neutron scattering, transport
   and magnetization measurements, of the interplay between
   superconducting (SC) and spin density wave (SDW) order in
   La$_2$CuO$_{4+y}$. Both kinds of order set in below the same
   critical temperature. However, the SDW order grows with applied
   magnetic field, whereas SC order is suppressed. Most importantly,
   the field dependence of the SDW Bragg peak intensity has a cusp at
   zero field, as predicted by a recent theory of competing SDW and SC
   order. This leads us to conclude that there is a repulsive
   coupling between the two order parameters. The question of whether
   the two kinds of order coexist or microscopically phase
   separate is discussed.
\end{abstract}


\maketitle

\section{Introduction}
The high-transition-temperature superconductors have dynamic and
sometimes static magnetic
order,\cite{YBCO,KeimerBSCCO,Yamada,Cheong,Tranquada,Kimura,YL}
whereas in conventional superconductors, magnetic and
superconducting order involving the same electrons typically do
not coexist. It is important to know whether these two kinds of
order compete or cooperate with one another, and whether they
coexist microscopically or form spatially separate phases. This
has been a subject of controversy concerning both experimental
results
\cite{YBCO,KeimerBSCCO,Yamada,Cheong,Tranquada,Kimura,YL,Savici,Uemura}
and theoretical predictions.
\cite{Zaanen,Kivelson,WhiteScalapino,Markiewicz,Martin} In the
context of interaction between SC and static SDW order,
excess-oxygen-doped La$_2$CuO$_{4+y}$ is especially interesting.
The transition temperature to long-range magnetic order ($T_m$)
coincides with the superconducting transition ($T_c$),~\cite{YL}
and both transition temperatures $T_c{ }\simeq T_m{ }\simeq 42$~K
are the highest achieved in La$_2$CuO$_{4}$ doped by any means, at
atmospheric pressure. Since $T_c$ and $T_m$ are so high, it has
been suggested that the SC and SDW order enhance one
another.\cite{YL} These observations place the oxygen-doped
material in strong contrast with La$_2$CuO$_{4}$ doped in
different ways, La$_{2-x-y}$Nd$_{y}$Sr$_x$CuO$_4$ \cite{Tranquada}
and La$_{1.88}$Sr$_{0.12}$CuO$_4$,~\cite{Kimura} in which SC and
SDW order appear to compete. This has raised the possibility that
the quenched disorder of Sr and Nd ions or the structural
distortion resulting from Nd substitution might obscure the
observation of the intrinsic interaction between SDW order and
superconductivity. By contrast, excess-oxygen-doped
La$_2$CuO$_{4+y}$ with $T_c{ }\simeq 42$~K is a stoichiometric
compound that exhibits three-dimensional order of the excess
oxygen, both parallel and perpendicular to the CuO$_2$
layers.\cite{YLunpublished}

We report here experimental results on this interesting compound,
which shed new light on the interaction of the SC and SDW orders.
We have measured the magnetic field dependence of the SDW Bragg
peak and find that the magnetic field enhances the SDW order while
resistivity measurements show that, as usual, the field suppresses
SC order. We also present detailed transport and magnetization
data. We synthesize our neutron scattering results with the
results of the muon spin rotation ($\mu$SR) experiments on the
same samples.~\cite{Savici,Uemura} These two experimental
techniques are complementary; the neutrons probe the long-distance
ordering of the Cu spins, while $\mu$SR measures the local
magnetic field due to the Cu spins at certain positions in the
unit cell.

Our paper is organized as follows: In Section II we provide
experimental details.  Section III is devoted to a presentation of
our results.  Finally, in Section IV we discuss the results and
draw conclusions.

\section{Experimental Details}
Single crystals of La$_2$CuO$_{4}$ have been grown by the
travelling solvent floating zone technique and subsequently
oxidized in an electrochemical cell, as described
previously.\cite{YL} It requires several weeks of electrochemical
oxidation to prepare a fully oxidized crystal with a volume of
order 1 cm$^3$. SQUID magnetization measurements have been made on
a small piece of each sample used for neutron measurements. The
magnetization studies evince a sharp single transition to the SC
state at $T_c =42$ K (onset), as well as the absence of weak
ferromagnetism, indicating an unobservable amount of remnant
undoped La${_2}$CuO$_4$. Thermogravimetric analysis has been
performed on two small single crystals oxidized to give $T_c=42$
K. We find oxygen concentrations of $y=0.10(1)$ and $y=0.12(1)$.
Since our large single crystals have the same $T_c$ we conclude
that their chemical composition is La${_2}$CuO$_{4.11}$. However
one cannot determine the hole density from this oxygen content.
Results of Ref.~\onlinecite{Chou} have shown that each
intercalated oxygen atom accepts two electrons at very low density
but accepts approximately one electron at higher density. Our
susceptibility measurements indicate that the hole density is $p
\simeq 0.14$, as discussed below.

Neutron measurements have been made on two different samples of
La$_2$CuO$_{4.11}$, Sample 1 and Sample 2, each of about 4.5 grams
in weight. They originate from different as-grown La$_2$CuO$_{4}$
crystals, but the preparation procedures are identical, and they
have the same $T_c$. Our previous report of SDW order and staging
behavior~\cite{YL} is based on measurements of Sample 2.
Magnetization measurements, as well as NMR and NQR studies by Imai
and coworkers~\cite{Imai} have also been made using pieces of
Sample 2.

\begin{figure}
\centering
\includegraphics[width=1.5in]{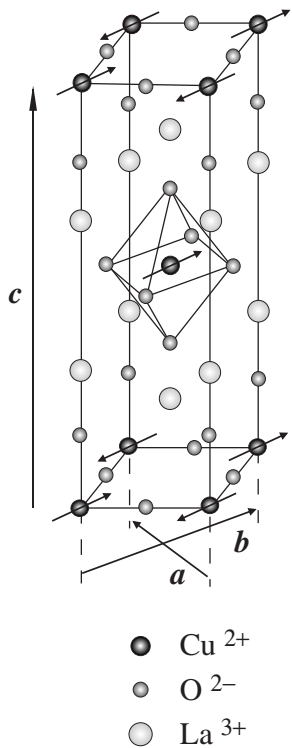}
\vspace{0in}%
\caption{\label{Structure}The unit cell of
La$_2$CuO$_{4}$, depicting the orthorhombic axis and spin
directions.}
\end{figure}

Transport measurements in a magnetic field have been made on
pieces of Sample 2, as well as on another small crystal, oxidized
to give $T_c=42$ K. We use the standard 4-probe technique on small
rectangular parallelepipeds, approximately $2\times1\times0.3$
mm$^3$ in size. Thin layers of Ag and, subsequently, Au are
evaporated onto the sample surface to form Ohmic contacts. The
current used for the four-probe measurements is 0.3 mA, below
which the resistivity is found to be current-independent.

Elastic neutron scattering studies were performed at the NIST
Center for Neutron Research in Gaithersburg, MD. We used the BT9
and BT2 thermal triple-axis spectrometers with incident neutron
energy of 14.7 meV, as well as the cold-neutron triple-axis
spectrometer SPINS with incident energy of 5 meV. A pyrolytic
graphite (PG) monochromator and PG analyzer were used, as well as
a PG filter to remove higher energy contamination from the
incident neutron beam. The magnetic field was applied using a 9 T
split-coil superconducting magnet.

\section{Results}
\subsection{Structural neutron scattering}
The structure of La$_2$CuO$_{4}$ is drawn in Figure
\ref{Structure}, showing the spin arrangement in the undoped
antiferromagnet. Because of a small tilt of the CuO$_6$ octahedra
relative to the {\em c}-axis, the crystal structure is actually
orthorhombic. We therefore use the orthorhombic unit cell with
{\em a} and {\em b} along the diagonals of the square and {\em c}
perpendicular to the layers.

Previous studies~\cite{YL,Wells} show that the excess oxygen in
La$_2$CuO$_{4+y}$ is not distributed uniformly, but rather has a
sine-density-wave modulation that is periodic along the {\em
c}-axis. The tilt angle of the CuO$_6$ octahedra changes sign
across the planes containing the most oxygen, so the tilt reversal
occurs every {\em
   n}th CuO$_2$ layer.  This behavior is called staging, and the sample
with tilt reversal every {\em n}th layer is called stage-{\em
   n}.~\cite{YL,Wells} The results reported here are for crystals that
are stage-4.

\begin{figure}
\centering
\includegraphics[width=2.5in]{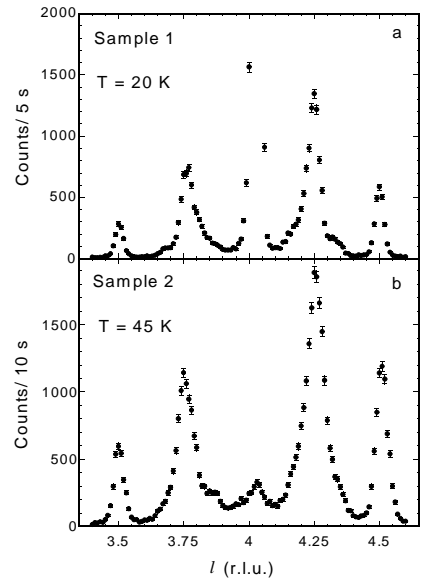}
\vspace{0in}%
\caption{\label{Stage_LowT}A scan along the {\em l}
    direction across the staging superlattice positions
    (0, 1, {\em l}) at low temperatures. (a) Sample 1, T = 20 K,
    $E_i=14.8$ meV, collimation 10'-40'-S-40'-open, 3-axis mode
    (S denotes the sample).  (b) Sample 2, T = 45 K, $E_i=35$ meV,
    collimation 20'-20'-S-20'-open, 2-axis mode.}
\end{figure}

Figure \ref{Stage_LowT} shows the intensities of neutron
scattering from nuclear Bragg peaks along the (0,1,{\it l})
direction. The undoped crystals have peaks at {\it l} even, and a
small residue of this can be seen at {\it l}$=4$ in each scan,
showing that a small component of the parent {\em Bmab} structure
remains in each crystal. There is also a very small amount of the
stage-6 compound. The two peaks displaced by $\pm 0.25$ reciprocal
lattice units (r.l.u.) around {\it l}$=4$ correspond to stage
four. Since the unit cell spans two CuO$_2$ layers, the position
1/4 corresponds to a period of eight layers. However, as discussed
previously,~\cite{Wells} the superlattice peak at 1/4 comes from
the ordering of the tilt angle of the CuO$_6$ octahedra, which has
an antiphase domain boundary every fourth layer, giving an overall
periodicity of eight layers.  Presumably, the antiphase domain
boundary allows more room for intercalated oxygen whose density
modulation has a period of four layers.

\begin{figure}
\centering
\includegraphics[width=2.5in]{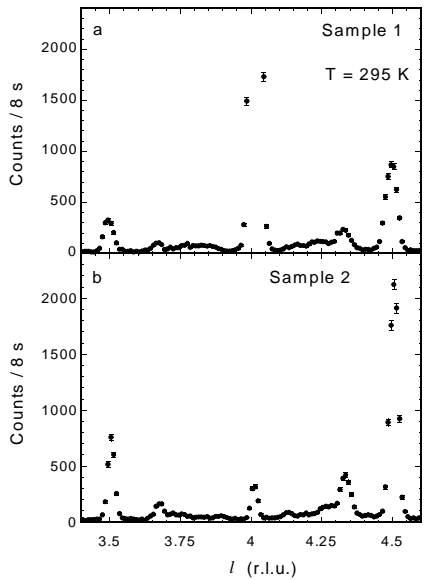}
\vspace{0in}%
\caption{\label{Stage_RoomT}A scan across the staging positions
(0, 1, {\em l}) at room temperature, T = 295 K. $E_i=5$ meV,
collimation 32'-40'-S-80'-open, 3-axis mode. (a) Sample 1, (b)
Sample 2.}
\end{figure}

The peak displaced by 0.5 r.l.u. might appear, at first sight, to
result from a stage-2 component. However, it most likely results,
instead, from the scattering from both the intercalated oxygen
itself and the concomitant displacements of the atoms around the
intercalants. Recent studies~\cite{YLunpublished} have shown that
in these samples the oxygen is three dimensionally ordered with a
period along the {\em c} axis of four layers, corresponding to a
wavevector displaced by 0.5 r.l.u. The three-dimensional ordering
of the intercalated oxygen persists up to 330 K, whereas the
stage-4 octahedral tilt ordering is lost above 295 K. Figure
\ref{Stage_RoomT} shows that the peaks displaced by 1/4 r.l.u. are
missing at room temperature, whereas those displaced by 1/2
remain.  When the crystal is {\em quenched} from above 330 K, the
three-dimensional ordering of the oxygen intercalants is lost and
the peak at 1/2 is not seen at low $T$, even though the octahedral
tilt ordering, evinced by the peak at 1/4, is
present.~\cite{YLunpublished} We conclude that the crystals are
stage-4 with very small inclusions of the oxygen-poor {\em Bmab}
phase as well as a small amount of stage 6. That the peaks at 1/4
and 1/2 come from a single phase is confirmed by observation that
the relative intensities of the these two peaks in Fig.
\ref{Stage_LowT} are almost the same for the two samples.

\subsection{Magnetization }

\begin{figure}
\centering
\includegraphics[width=3in]{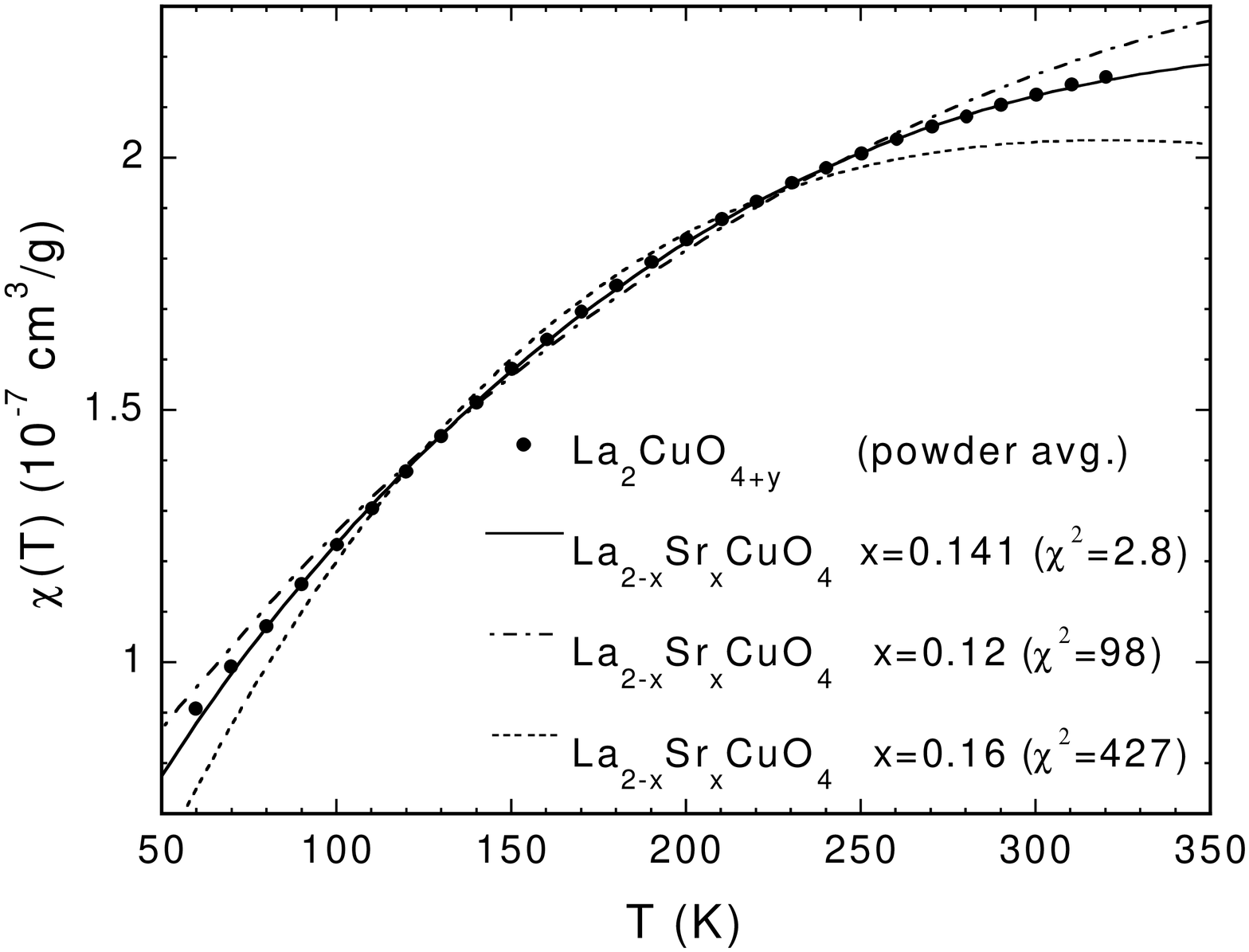}
\vspace{0in}%
\caption{\label{Chi}Temperature dependence of $\chi$
     in La$_2$CuO$_{4+y}$ (Sample 2) compared with
     La$_{2-x}$Sr$_x$CuO$_4$. One set of measurements was taken
     with the field alignment $H\|c$ and another set was taken with
     $H\|ab$. The powder-averaged susceptibility $\chi(T)$ was
     calculated from our data via
     $\chi(T)=\frac{1}{3}\chi^c+\frac{2}{3}\chi^{ab}$.
     The curves correspond to the scaling function $F$ with
     parameters obtained from La$_{2-x}$Sr$_x$CuO$_4$ for
     different doping levels {\em x}. We extract the functional form
     of $F$ from Ref. \onlinecite{Nakano}, a study
     of polycrystalline La$_{2-x}$Sr$_x$CuO$_4$.
     The ``$\chi^2$'' goodness-of-fit (not to be confused with
     susceptibility) is displayed next to each fit.}
\end{figure}
We next discuss measurements of the uniform magnetic
susceptibility $\chi(T)$ in our excess oxygen-doped samples and
their relationship to similar measurements in the Sr-doped
material La$_{2-x}$Sr$_x$CuO$_4$, which allow us to estimate the
hole concentration. The normal state susceptibility of
La$_{2-x}$Sr$_x$CuO$_4$ depends on the temperature $T$ and hole
concentration $p$ according to
$\chi(p,T)=\chi_0(p)+\chi^{2D}(p,T)$, where $\chi_0$ is a
temperature-independent constant. As shown in Refs.
\onlinecite{Johnston} and \onlinecite{Nakano}, the
temperature-dependent term $\chi(T)$ follows a scaling form
$\chi^{2D}(p,T/T_{max})/\chi^{2D}_{max}(p) = F(T/T_{max}(p))$,
where {\em F} is the scaling function. This scaling is found to
describe the data for La$_{2-x}$Sr$_x$CuO$_4$ for a wide range of
{\em p}, up to $p=0.26$.\cite{Johnston,Nakano} The powder-averaged
susceptibility of our La$_2$CuO$_{4.11}$ is shown in Fig.
\ref{Chi}. The three lines are fits using the scaling function $F$
with the parameters $\chi^{2D}_{max}$ and $T_{max}$ chosen to
match those in La$_{2-x}$Sr$_x$CuO$_4$ with $x=0.141, 0.12$, and
0.16, respectively. We find that the curve corresponding to
$p=0.141$ best describes the data, suggesting that the hole
concentration for stage-4 La$_2$CuO$_{4.11}$ is $p\simeq 0.14$.
The fact that this functional form fits both
La$_{2-x}$Sr$_x$CuO$_4$ and La$_{2}$CuO$_{4.11}$ suggests that the
hole homogeneity is similar for the two materials, at least for
temperatures above $T_c$.  Thus the hole concentration appears to
be quite uniform even though the oxygen concentration varies
periodically along the {\em c} axis.

The undoped component of these crystals is fractionally very
small, as we conclude from the following observations: The weakly
doped antiferromagnet reveals itself as a peak in the magnetic
susceptibility at about 260 K, which arises from hidden weak
ferromagnetism. At fields high enough to induce the weak
ferromagnetic transition, the moment induced can be used to
measure the volume of antiferromagnet in the
sample.\cite{WellsZPhys} SQUID magnetization measurements at high
fields show that this signature of antiferromagnetic order becomes
unobservable as a result of electro-chemical doping, implying that
the antiferromagnetic inclusions in our crystals correspond to at
most a few per cent of their volume. Similarly, we can use the
intensity of inelastic scattering near the antiferromagnetic Bragg
peak, resulting from spin waves, to measure the undoped
fraction.~\cite{YL} No commensurate component is detectable in
inelastic scans, confirming that the antiferromagnetic fraction is
very small. These observations are equally true for both Sample 1
and Sample 2.

\subsection{Elastic magnetic scattering}
\begin{figure}
\centering
\includegraphics[width=3in]{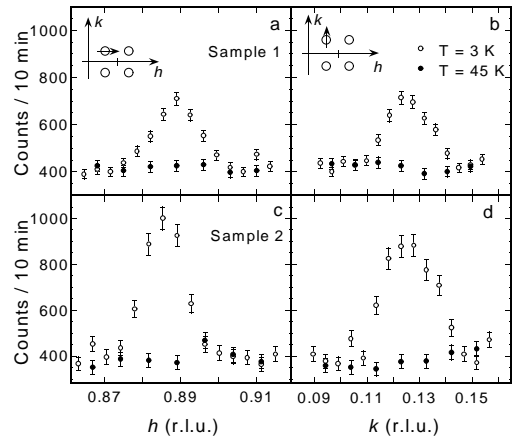}
\vspace{0in} %
\caption{\label{Bragg}Neutron elastic scattering from a SDW Bragg
    peak at zero applied field. The sample was oriented such that
    the neutron wave vector transfer Q was parallel to the {\it
    ab}, Cu-oxygen, planes. The reciprocal lattice position is Q =
    (0.89, 0.128, 0). (a)and (b) are scans for Sample 1 along
    reciprocal space directions {\em h} and {\em k}, respectively.
    (c) and (d) are scans for Sample 2 along {\em h} and {\em k},
    respectively. The insets show schematically the
    reciprocal space and the scan directions. The temperatures are
    3 K (open circles) and 45 K (dots) and the collimation is
    32'-40'-20'-open, $E_i=5$ meV.}
\end{figure}

As reported previously, the staged compound exhibits two
dimensional static SDW order with a periodicity in the square
CuO$_2$ layers that is approximately 8 times that of the
underlying lattice, consistent with the stripe model of Tranquada
{\it el al}.~\cite{Tranquada} This manifests itself in elastic
neutron scattering as four peaks around the 2D antiferromagnetic
zone centers (1,0,{\em l}) and (0,1,{\em l}). (Recall that we are
using an orthorhombic unit cell.) Thus, the first quartet of SDW
Bragg peaks is at Q = ($1\pm\delta_h, \pm\delta_k, 0$). (See the
insets of Fig. \ref{Bragg}). For Sr doped La$_2$CuO$_{4}$, the
incommensurability ($\delta_{h,k}$) depends on doping, but for the
stage-4 compound we always find the values $\delta_h$ = 0.114 and
$\delta_k$ = 0.128 r.l.u., corresponding to a shift of $0.121(2)$
r.l.u.  for the tetragonal unit cell, close to the value for Sr
concentration 1/8 per Cu atom.\cite{Kimura}

Figures \ref{Bragg}(a,b) show the SDW peaks at (0.89,0.128,0) for
Sample 1; similar data are presented in Fig. \ref{Bragg}(c,d) for
Sample 2 with the identical spectrometer configuration, so a
direct comparison between the two samples can be made. Scans along
two perpendicular directions in reciprocal space, {\em h} and {\em
k}, are shown for each sample. Comparing the two samples we see
that the peak positions are identical, and the widths are similar,
but the intensity of the signal for sample 2 is about twice as
large as that for sample 1.

The full width at half maximum of the peak, 0.008 r.l.u., is
resolution limited for this spectrometer configuration, which
indicates that the SDW order has a correlation length greater than
600 $\AA$. Previous studies of the stage-4 crystal have indicated
that the SDW is ordered over distances $\ge 600$ $\AA$ within the
CuO$_2$ plane and there is short-range order ($\sim 13$ $\AA$)
perpendicular to the planes. The three-dimensional structure is
consistent with a collinear SDW, with the {\em local} spin
structure identical to that of the undoped insulator.~\cite{YL}

\begin{figure}
\centering
\includegraphics[width=3in]{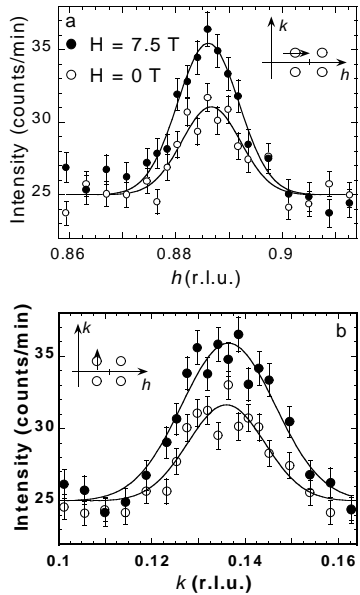}
\vspace{0in}%
\caption{\label{Bragg_H}SDW peaks with and without an applied
    field at the same spectrometer configuration as for Fig.
    \ref{Bragg}. The open circles are for zero field and the filled
    circles are for H = 7.5 T. The lines are the results of
    Gaussian fits. The inset shows schematically the position and
    scan direction in reciprocal space. The field is applied
    parallel to the {\em c}-axis, perpendicular to the planes. The
    Bragg peak shown occurs at {\bf Q} = (0.886, 0.132, 0) and the
    temperature is T = 1.5 K. The collimation is 32'-40'-20'-open,
    $E_i=5$ meV. These results are for Sample 1 with scans along
    (a) {\em h} and (b) {\em k}.}
\end{figure}

Our most surprising results involve the magnetic field dependence
of the SDW Bragg peak.  Figure \ref{Bragg_H} shows typical scans
across one of the incommensurate peaks for Sample 1 with and
without a magnetic field applied parallel to the {\em c} axis.
When the sample is cooled in an applied field of 7.5 T, the
scattering intensity grows dramatically.  Scans along both {\em h}
and {\em k} demonstrate that the peak position and width remain
the same within the errors when the field is applied. The same
relative increase in intensity is observed at incommensurate
positions near different reciprocal lattice points, using
different neutron energies (5 and 14.7 meV) and on different
spectrometers. Sample 2 shows similar results, albeit with a
smaller increase in the peak intensity, as discussed further
below. Because the sample is mounted in the split-coil magnet to
allow application of the field in the {\em c} direction, we are
able to measure scattering only in the {\em {ab}} plane and can
not scan the momentum along {\em c}.

The fact that the increased scattering at high fields is at
exactly the same position in reciprocal space as that at zero
field puts strong constraints on any possible theoretical models.
This is especially striking because the peak occurs at a position
that is incommensurate in both the {\em a} and {\em b} directions.
Specifically, the increase must reflect an enhancement or a
proliferation of the existing SDW order rather than the creation
of a new SDW state, for example, in the vortex cores.

\subsection{Onset of SDW and SC order}

\begin{figure}
\centering
\includegraphics[width=3.5in]{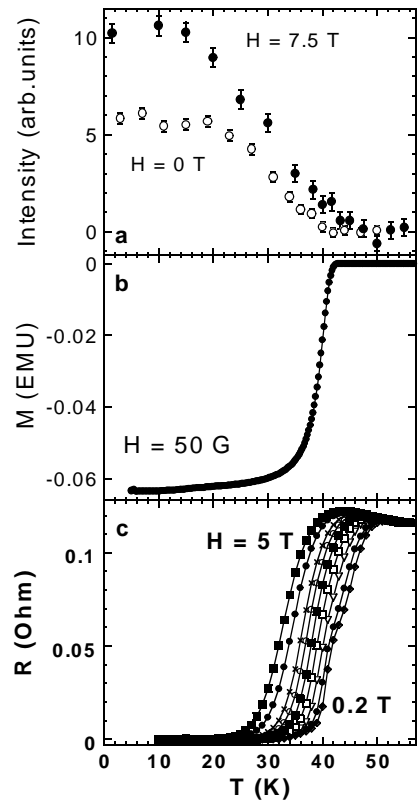}
\vspace{0in}%
\caption{\label{SDW_T}Temperature dependence of the
Bragg peak
     intensity, superconducting shielding and in-plane resistivity.
     (a)The peak intensity for Sample 1 at zero field (open circles)
     and at 7.5 T (filled circles). The peak position, collimation
     and energy
     are the same as for Fig. \ref{Bragg}. The points are calculated
     by averaging longitudinal and transverse scans around the peak
     position; a constant background has been measured and
     subtracted. (b) The magnetization at small
     field showing the sharp onset of superconductivity at the same
     temperature at which the SDW order sets in.  (c) The in-plane
     resistance vs. temperature at applied fields of 0.2, 0.5, 1,
     1.5, 2, 2.5, 3, 4, and 5 T. The crystal was oriented by a Laue
     diffractometer in order to ensure that the transport current
     was parallel to the {\em ab}-planes, while the magnetic field
     was applied parallel to the {\em c}-axis, as in the neutron
     scattering experiments.}
\end{figure}

Figure \ref{SDW_T} shows the SDW peak intensity (a) and the
resistivity (c) at different applied fields as functions of
temperature. It is clear that the applied magnetic field results
in an increase of the amplitude of the magnetic Bragg peak over
the entire range of temperatures, whereas there is, at most, a
very small increase in the SDW ordering temperature. On the other
hand, the SC transition temperature, as determined by the onset
$T_R$ of the decrease in resistivity, shifts substantially toward
lower temperature with increasing field. Similar behavior of the
resistivity in a magnetic field has been reported for Sr-doped
La$_{2-x}$Sr$_{x}$CuO$_4$ at similar doping levels.\cite{Katano}
Clearly, $T_{R}\simeq 50$ K in zero field, is considerably higher
than $T_{c}\simeq 42$ K measured by the onset of diamagnetism
(Fig. \ref{SDW_T}b). This is possibly the result of SC filaments
that form at higher $T$ and shunt the transport current but do not
give rise to a large diamagnetic signal. The SDW onset at
$T_{m}\simeq 42$ K coincides with the magnetically determined
$T_c$.

These results show that the coincidence between $T_c$ and $T_m$
occurs only for $H = 0$, which thus becomes a special point in the
phase diagram. A similar coincidence between $T_c$ and $T_m$ at
zero field has recently been observed in more lightly doped
La$_2$CuO$_{4+y}$ (stage 6) with $T_c$ = 34
K.~\cite{YLunpublished}

The transport measurements in Fig. \ref{SDW_T} have been made with
the current and voltage contacts placed on the top surface of a
crystal. This makes it difficult to determine even nominal
resistivity, given the high electrical anisotropy of
La$_2$CuO$_{4+y}$. Measurements on a different sample, with
contacts positioned on the side surfaces, have resulted in
$\rho_{ab}(0)\simeq 2$ m$\Omega$-cm, which is an order of
magnitude higher than that of the best optimally-doped samples of
La$_{2-x}$Sr$_x$CuO$_4$. The high value of resistivity is similar
to that reported in Ref.~\onlinecite{Katano} for
La$_{1.88}$Sr$_{0.12}$CuO$_4$ (1/8 Sr-doping), which shows a
static SDW similar to that in our samples. This suggests that the
large resistivity may be related to the incipient SDW order.
However, our samples often contain a number of small cracks that
appear to grow during the electro-chemical doping. This makes it
impossible to accurately determine the current density, so
conclusions from the high resistivity must be drawn with caution.

\subsection{Field dependence of the elastic magnetic scattering}
\begin{figure}
\centering
\includegraphics[width=3in]{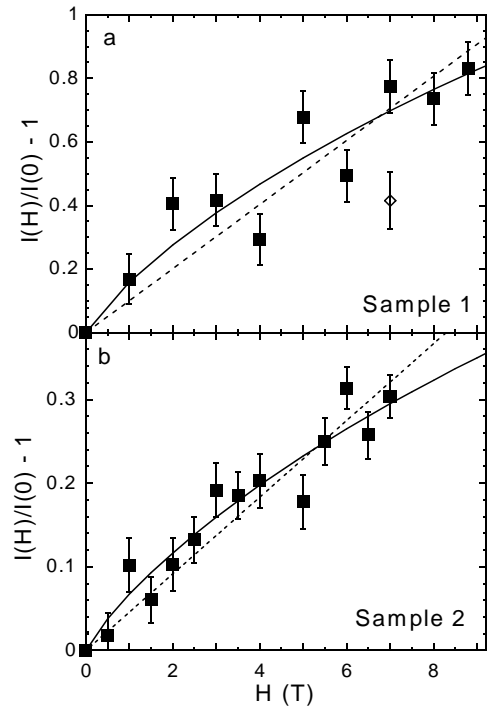}
\vspace{0in}
\caption{\label{SDW_H} Relative change in SDW peak intensity above
     background vs. field for (a) Sample 1 and (b) Sample 2.
     The filled squares are for measurements made after cooling in a
     field (FC). The initial neutron energy is $E_i = 14.7$
     meV (BT-2 spectrometer). The background at 50 K (found to be
     independent of field) has been subtracted and the difference
     is then normalized to the zero-field value. The dashed line
     assumes $\Delta I \sim |H|$, whereas the solid line
     corresponds to the function given in ref. \onlinecite{Demler},
     $\Delta I \sim H/H_{c2}\ln(3H_{c2}/H)$. For the latter we have
     fixed
     H$_{c2}$ at 60 T, so that the only free parameter is the
     prefactor. We have also performed a zero-field cooled (ZFC)
     measurement (diamond in the figure), in which a field of 7 T
     is applied after the sample is first cooled to 1.5 K.}
\end{figure}

In order to elucidate the origin of the increase of the SDW peak
amplitude, we have carried out measurements of the intensity of
the elastic neutron scattering as a function of field. The results
for both samples are presented in Figure \ref{SDW_H}.
Surprisingly, we find that the increase of the intensity with
field up to $H = 9$ T is approximately proportional to $|H|$
rather than $H^2$. This makes $H = 0$ a singular point; that is,
$I(H)$ is not analytic at $H = 0$. We have confirmed that the
scattering intensity depends only on the magnitude and not on the
sign of the field.

As is clear from Fig. \ref{SDW_H}, the relative increase of the
SDW peak intensity is about twice as large for Sample 1 as it is
for Sample 2.  Since the SDW peak intensity at zero field is about
twice as large for Sample 2 as for sample 1, the absolute
intensity change is similar for the two crystals. The SDW
intensity has been normalized to the mass of the crystal for each
sample. We have checked that the intensities of the nuclear Bragg
peaks are consistent with this normalization.

Enhancement of SDW long-range order by a magnetic field has also
been observed in Sr-doped La$_{2-x}$Sr$_x$CuO$_4$, $x \approx
1/8$.\cite{Katano} The authors of Ref. \onlinecite{Katano} have
not determined the field dependence of the peak intensity. Related
effects have been observed in optimally and slightly under-doped
La$_{2-x}$Sr$_x$CuO$_4$.~\cite{Lake,Lake2}

As expected, the zero-field-cooled (ZFC) intensity (the point at 7
T for Sample 1) is reduced significantly, compared to the
field-cooled (FC) value, but it is still higher than the
zero-field value. The magnetization of La$_2$CuO$_{4+y}$ is
irreversible at 1.5 K and 7 T because of the pinning of vortices.
When the superconductor is cooled in a field, pinned magnetic flux
lines permeate the bulk of the sample, and the magnetic field is
expected to be distributed homogeneously. This is because the
London penetration length $\lambda \simeq 2200$ $\AA$ is much
larger than the inter-vortex distance of 150 $\AA$ at 7 T.
However, when the field is applied after the superconductor is
cooled to 1.5 K, the magnetic field penetrates only partially into
the bulk of the sample since strong pinning prevents vortices from
moving freely. Therefore, the bulk of the sample is better
shielded from the magnetic field after ZFC and the SDW peak
amplitude is therefore smaller than after FC.

\section{Discussion and Conclusions}

Since we observe both SC and SDW order it is natural to ask
whether these occur in a single phase or in separated phases. In
particular, we must consider the possibility that phase separation
results from inhomogeneity in the hole concentration.  As
discussed above, the magnetic susceptibility suggests a uniform
hole density. Additional information about variations in the hole
concentration comes from a detailed NMR and NQR study carried out
by Imai and coworkers on one of our La$_2$CuO$_{4.11}$
samples.\cite{Imai} Interestingly, the $^{63,65}$Cu NQR line
profiles and positions turn out to be nearly identical in
La$_2$CuO$_{4.11}$ and La$_{1.865}$Sr$_{0.135}$CuO$_4$. This
indicates, first, that the two materials have comparable hole
concentrations and second, that the distribution of electric field
gradients at the Cu sites in the two materials is the same
implying that they have comparable homogeneities. This also
necessitates that in La$_2$CuO$_{4.11}$ there is no measurable
difference in hole concentrations for the different CuO$_2$
layers.

One of the best measures for the homogeneity of
La$_2$CuO$_{4}$-based systems is the sharpness of the
tetragonal-to-orthorhombic phase transition. From previous studies
of this transition we know that crystals of
La$_{2-x}$Sr$_{x}$CuO$_{4}$, grown by the travelling solvent
floating zone technique, have quite uniform Sr distributions.
Technically, the sharpness of the structural phase transitions
implies that any correlations of the local phase transition
temperatures at distance $r$, due to a tendency of the Sr dopants
to cluster, must fall off faster than $r^3$.~\cite{Halperin} Of
course, there is local disorder caused by the Sr$^{2+}$ dopants,
but this is absolutely statistical so that on large length scales
the material is homogeneous. The fact that La$_2$CuO$_{4.11}$ has
an identical $^{63,65}$Cu NQR line profile to that in
La$_{1.865}$Sr$_{0.135}$CuO$_4$ requires that it be similarly
homogeneous.

Inelastic neutron scattering results also give strong evidence for
the homogeneity of the doping in our crystals. The width of the
incommensurate inelastic peaks, reported in Ref.~\onlinecite{YL},
is as small as that of the narrowest peaks observed in
La$_{2-x}$Sr$_x$CuO$_4$. Since the incommensurability varies
continuously with the doping in
La$_{2-x}$Sr$_x$CuO$_4$,\cite{Yamada} an inhomogeneous hole
concentration should result in an increase in the width of the
inelastic peaks. Furthermore, the width at $x=0.125$ is
significantly smaller than that at other {\em x}'s, so our narrow
inelastic peak indicates that the hole concentration is quite
uniform.

Our field-enhanced magnetic scattering has interesting
implications. The singular dependence of the SDW amplitude on
applied field excludes a purely magnetic mechanism for the
phenomenon such as a suppression of fluctuations of the ordered
moment by the applied field, as suggested by Katano {\it et
al.}~\cite{Katano}, or an increase in the correlations along the
c-axis. The intensity of a magnetic Bragg peak $I(H)$ is
proportional to the square of the ordered staggered moment,
$I(H)\sim |M^\dagger|^2$. By symmetry, the first non-zero
correction to $M^\dagger$ must be of second order in field:
$\Delta M^\dagger(H)\sim H^2$. Therefore, the leading correction
to the intensity from a purely magnetic mechanism must be
quadratic in the field $\Delta I(H)\sim H^2$, rather than $\Delta
I(H)\sim |H|$, as found experimentally.

The linear increase of the magnetic signal with $|H|$ suggests
that the effect originates from magnetic flux lines penetrating
the sample in the superconducting state. Indeed, since every
vortex carries one magnetic flux quantum, the number of vortices
is proportional to $|H|$. The role of vortices has been emphasized
in Reference \onlinecite{Katano} and in a closely related
experiment on inelastic and elastic neutron scattering from
optimally doped La$_{2-x}$Sr$_x$CuO$_4$ (x=0.163) in a magnetic
field.~\cite{Lake} The latter experiment has revealed an increase
with field of the intensity of low (but non-zero) energy spin
excitations at the incommensurate positions.
The low-energy fluctuations exist in the normal state, but they
are suppressed below $T_{c}$ in zero field because of the opening
of a gap for spin excitations. The authors of reference
\onlinecite{Lake} ascribe the increase of the subgap fluctuations
below $T_c$ with applied field to fluctuations toward magnetic
order in the vortex cores. A very recent paper by B. Lake
{\em et al.}\cite{Lake2} discusses neutron measurements on
La$_{1.9}$Sr$_{0.1}$CuO$_4$.  Lake {\it et al.} observe results
that are consistent with ours, despite the fact
that the SDW order is known to be of shorter range in
La$_{1.9}$Sr$_{0.1}$CuO$_4$ than
in La$_2$CuO$_{4.11}$.~\cite{Yamada}

A recent theoretical model~\cite{Demler,Zhang} for homogeneous
coexistence of SDW and SC order explains the singular $\sim |H|$
dependence of the intensity in a natural way. In this model the
SDW order parameter $|\phi|^2$ is directly coupled via a positive
(repulsive \cite{Zhang}) coupling coefficient to the
superconducting order parameter $|\psi|^2$. The singular increase
of the peak intensity results from the singular response of a
superconductor to an applied magnetic field. In the presence of
vortices, the SC order parameter is reduced in the entire volume
of the superconductor, even outside the vortex
cores,~\cite{Brandt} thus causing an increase in the magnetic
order parameter because of the positive coupling between
$|\phi|^2$ and $|\psi|^2$. Since the SDW correlation length spans
many inter-vortex distances, $|\psi|^2$ is averaged over the
regions outside the cores. As a result, $|\phi|^2$ is proportional
to the absolute value of magnetic field, with a logarithmic
correction. This, in turn, leads to the approximately linear
increase of the SDW peak intensity at small fields and non-zero
derivative $dI/dH$ at $H =0$. We show in Fig. \ref{SDW_H} the form
$\Delta I \sim H/H_{c2}\ln(3H_{c2}/H)$ predicted by
ref.~\onlinecite{Demler}, which describes our data quite well. The
fit has only one adjustable parameter since $H_{c2}$ is fixed at
60 T.\cite{Ando}

Measurements of muon spin relaxation ($\mu$SR) have been made on
our Sample 2.~\cite{Savici,Uemura} At temperatures well below
$T_c$ the muons' spins precess in an internal magnetic field
distribution, which appears to be very similar to that resulting
from the SDW in La$_{1.47}$Nd$_{0.4}$Sr$_{0.13}$CuO$_4$. However,
the signal from Sample 2 corresponds to only a fraction,
$\sim40$\%, of the muons experiencing the
field.~\cite{Savici,Uemura} This suggests that the system has
spatially separated into magnetically (SDW) ordered and
non-magnetic phases. The average precession frequency reaches its
low-temperature value quite abruptly below the SC $T_c$ in
La$_2$CuO$_{4.11}$, whereas it increases gradually below the SDW
transition in La$_{1.47}$Nd$_{0.4}$Sr$_{0.13}$CuO$_4$.  This
suggests that putative microscopic phase separation occurs at
$T_c$.~\cite{KivelsonPNAS} Modelling of the $\mu$SR
signal~\cite{Savici,Uemura} suggests that the typical size of the
magnetic regions is of order 15-50 $\AA$. Note that $\mu$SR
measurements are strongly suggestive but not conclusive about the
volume fraction and size of the superconducting regions.

This model~\cite{Uemura} of microscopic phase separation may help
explain the difference in the field dependence between Sample 1
and Sample 2 (see Fig. \ref{SDW_H}). We assume that the flux
penetrates non-magnetic and SDW regions at random, but the field
enhances SDW order primarily in the non-magnetic phase.  The
field-induced change in $F$, the fraction of the sample primarily
in the SDW phase, is proportional to the fraction $1-F$ primarily
in the non-magnetic phase. Therefore, the relative change of the
Bragg peak with field, at small field, is expected to be
proportional to $(1-F)/F$. The muon measurements tell us that for
Sample 2, $F_2\sim0.4$. Since the Bragg peak intensity for Sample
2 is twice that of Sample 1, we would expect $F_1\sim0.2$ and the
relative change with field would be about 2.5 times smaller for
Sample 2 than for Sample 1, in agreement with observation.
Essential to this argument is the experimental fact that the
detailed geometry of the SDW scattering at high fields is
identical to that at zero field.

Since the non-magnetic (or weakly magnetic) regions are
superconducting, we propose that the enhancement of SDW order is
the result of the Demler {\it et al.}~\cite{Demler} mechanism
discussed above. Of course, the magnetic regions may also be
superconducting, but since the SDW order parameter in them must be
large, flux penetration cannot enhance it significantly.

If, as suggested by $\mu$SR, 40\% or less of each layer of Sample
2 consists of SDW ordered regions that are only 15-50 $\AA$ in
size, it is at first surprising that long range SDW order is
observed, even in zero field. One might have ascribed this to
simple nearest-neighbor percolation, since 40\% is close to the
percolation threshold in two dimensions. However, Sample 1 has
only half the SDW concentration, judging from the size of the
Bragg peak, and it also shows long-range order. Furthermore, muon
experiments on a crystal of La$_{1.88}$Sr$_{0.12}$CuO$_4$ show
only $\sim12$\% SDW volume fraction and such crystals also show
long range magnetic correlations.~\cite{Kimura} Thus, if the
system is microscopically phase separated into regions of typical
size 15 $\AA$ - 50 $\AA$, there must be coupling between SDW
regions via the intervening non-magnetic SC region. We suggest
that the magnetic order parameter has a long tail inside the SC
region with a magnetic moment that is below the $\mu$SR
resolution.~\cite{palee_private} This would give rise to
percolation of the long-range order. We note that neutron
scattering and $\mu$SR have different time scales, which must be
taken into account in comparing the results.~\cite{Uemura}
However, as demonstrated in Reference~\onlinecite{Uemura}, despite
their different energy resolutions $\mu$SR and neutron scattering
measure exactly the same temperature dependence for the order
parameter implying that they are seeing identical physics.

Our observation of competing SC and SDW order is unexpected in
light of the coincident transition temperatures to SC and magnetic
order. One possible explanation of this is that the transition
occurs at a tetracritical point.\cite{Zhang,Aharony,KivelsonPNAS}

We conclude that there is an interesting interplay between the SDW
and SC orders in the high $T_c$ superconductors. Although the two
kinds of order appear at the same temperature, there is clearly a
competition between them. The muon experiments together with the
sample variation we see suggest that microscopic electronic phase
separation may occur when superconductivity sets in. The recent
theory of Demler and Sachdev,~\cite{Demler} based on the
assumption of the microscopic coexistence and competition between
SC and SDW order parameters explains the field dependence of the
SDW Bragg peak.

\begin{acknowledgments}
We have benefited from useful discussions with E. Demler, S.
Sachdev, S.-C. Zhang, P. A. Lee and Y. J. Uemura. We thank T. Imai
for discussions of the NMR and NQR results and for permission to
summarize these results before publication. This work has been
supported at MIT by the MRSEC Program of the National Science
foundation under Award No. DMR 9808941, by NSF under Awards No.
DMR 0071256 and DMR 99-71264. Work at the University of Toronto is
part of the Canadian Institute for Advanced Research and is
supported by the Natural Science and Engineering Research Council
of Canada. We acknowledge the support of the National Institute of
Standards and Technology, U.S. Department of Commerce, in
providing the neutron facilities used in this work. Work at SPINS
is based upon activities supported by the National Science
Foundation under Agreement No. DMR-9986442.
\end{acknowledgments}


\end{document}